\documentclass[]{article}
\input psfig.sty

\begin{document}
\begin{center}
\Large VISUAL PHOTOMETRY: COLOUR AND BRIGHTNESS SPACING OF COMPARISON STARS

\vspace{1.0cm}

\normalsize
{\it by Alan B. Whiting \\
University of Birmingham} 
\end{center}

\vspace{1.0cm}

\large
A significant amount of data on the historical and current behaviour
of variable stars is derived from visual estimates of brightness using
a set of comparison stars.  To make optimum use of this invaluable
collection one must understand the characteristics of visual photometry,
which are significantly different from those of electronic
or photographic data.  Here I show that the dispersion
of estimates among observers is very consistent at between 0.2 and 0.3
magnitudes and, surprisingly, has no apparent dependence on the colour
of comparison stars or on their spacing in brightness.
\normalsize

\vspace{1.0cm}

\noindent {\it Introduction: Visual Photometry}

Measuring the brightness of an object is one of the most basic of
operations in astronomy and doing it accurately is one of the most
important.  For most of the history of the science it has been accomplished
by the human eye, supplemented for about the past century by 
photographic and electronic methods.  These are certainly more accurate
and objective than visual estimates, but they have not entirely
supplanted the eye for two major reasons.  First, many
historical records are only
visual in nature; the best CCD in the world today cannot measure Eta Carinae
in 1835.  Second, even in this age of large-scale, rapid surveys, visual
estimates may be the only way of getting the desired temporal coverage.
Observers from the American Association of Variable Star Observers
(AAVSO\footnote{www.aavso.org}), for example,
are routinely called upon to alert professional astronomers to
outbursts or other behaviour of objects in support of observing campaigns.
In this way an amateur with a small telescope may trigger the use of
the Hubble Space Telescope and visual data may be combined with far more
sophisticated measurements.  As a recent example,
Humphreys et al.$^1$ used visual estimates of Eta Carinae in conjunction with
instrumental photometry and spectroscopy from a variety of telescopes.

But the human eye is not a simple detection and measuring system.  In
a sense, there is no such thing as raw visual photometric data; everything
is heavily processed before it can be recorded.  While this automatic processing
is no doubt useful in a terrestrial environment it is occasionally
annoying to the photometrist.  Known effects include the tendency for red
stars to appear brighter if stared at for a time and for a star placed
vertically above a similar one in the visual field
to appear brighter.  And, even more than instruments,
people differ among themselves.

To better understand the workings of the visual photometry system here I
concentrate on two features of comparison stars, motivated by experiences
of my own in making visual estimates. 

All the data analysed were produced using the current visual method
as practised by the AAVSO and similar organisations.  The observer is
provided a chart of suitable scale, centred on the variable star of interest.
On it are marked comparison stars whose $V$ magnitude has been measured by
electronic means, the latter given to one decimal place.  The observer 
picks stars brighter and fainter than the variable and then judges whether it
is closer to one than the other in brightness, and by how much; 
converts this to an estimated magnitude; then reports
that estimate.  While some exceptional observers can reliably
see smaller differences,
all data are reported to a tenth of a magnitude.  Photometry of comparison
stars in other bands is also readily
available, mostly for the use of CCD observers.

The first feature we investigate is best illustrated by an anecdote.  
The author, AAVSO
chart in hand, was making an estimate of Eta Carinae standing next to a student.
The latter, using the same comparison sequence at the same time, was firm
in giving an estimate of 5.1; the author, equally firm at 5.0.  Neither
could convince the other.  A plausible explanation comes from the fact that
the lens of the eye yellows with age, so the student was seeing more blue
light than the (much older) author, and thus made Eta with its strong H-alpha
emission relatively less bright.  The first investigation seeks the effect
of the colour of the comparison stars on the spread of visual estimates.
The hypothesis to be tested holds that the dispersion of estimates will
increase as the colour of the comparison stars differs from that of the
variable.

The second investigation stems from the perceived difficulty in
placing a variable when there is a great difference in brightness between
the bracketing comparison stars.  The unreliability of the human eye when
there are no nearby guides is suggested also by a table given by Webb$^2$,
comparing the magnitudes assigned to telescopic stars by four outstanding
visual observers of the nineteenth century.  Extrapolating the generally
agreed system of six naked-eye magnitudes, the four (F. W. Argelander, F. R. W.
Struve, Adm. W. H. Smyth and Sir John Herschel) disagreed at the
half-magnitude level by about 6.5, sometimes by a full magnitude at 8.5
and worse at fainter levels.  The second investigation therefore seeks
how much disagreement in visual estimate can be traced to large differences
in brightness between comparison stars.  The hypothesis to be tested
asserts that the dispersion among estimates will increase as the difference
in brightness between the bracketing comparison stars increases.

\vspace{0.5cm}
\noindent {\it The Sample and Processing}

While there are several organisations worldwide of dedicated variable-star
observers, the largest and the one with the most accessible data is the
American Association of Variable Star Observers, whose data were used
exclusively for this study.

For the investigation we need variable
stars with many observations (to give a good
delineation of the spread of estimates) and a variety of comparison stars. 
The
latter translates directly into a large variation in brightness.  These
conditions
essentially limit us to bright Mira-type variables.  We must then bear in
mind that what we find could, in principle, be different for a different
type of star; but it immediately simplifies the colour analysis, since
essentially no comparison stars will be redder than the variable.

Nine stars were chosen from the AAVSO list fitting the requirements and
I downloaded from the website one to two periods of visual observations 
each$^3$.  For each observation
the Julian date and time; the reported brightness; 
and the brightness of the two reported comparison stars were extracted.  

I initially tried to fit a polynomial to produce a smooth curve for
reference (as did Price, Foster \& Skiff$^4$), but found no number of terms
that would reproduce well the overall form without adding artifacts due
to unfortunate or badly-placed odd observations.  In the end I used
a top hat smoothed version, its width depending on the density of 
observations.  For each observation (not the corresponding
magnitude on the smoothed curve) I determined, from AAVSO photometry,
the $B-V$ colour of the comparison star nearest in brightness.  (The
$V-R$ colour might have been more directly applicable to visual
observations of Miras, but
was unavailable for many stars.)  Also for each observation, I
recorded the difference in magnitude between the bracketing comparison stars.

In subsequent displays time is presented in days from the first observation
used.  Table~\ref{observations} gives the starting Julian Date and number
of observations for each star analysed, to allow connections with other
investigations and to give an idea of the quantity of data available.

\begin{table}
\begin{tabular}{lcc}
Star & Starting JD & No. of Observations \\
R Leonis & 2455472.0 & 756 \\
R Aquarii & 2455032.8 & 148 \\
R Bootis & 2455231.5 & 622 \\
R Canum Venaticorum & 2455223.5 & 319 \\
R Hydrae & 2455170.8 & 67 \\
S Coronae Borealis & 2455301.2 & 407 \\
T Ursae Minoris & 2455400.5 & 414 \\
R Ursae Majoris & 2455404.5 & 344 \\
T Ursae Majoris & 2455400.2 & 468 \\
\end{tabular}
\caption{Summary observational data on the stars analysed.}
\label{observations}
\end{table}

\begin{figure}
\centerline{\psfig{file=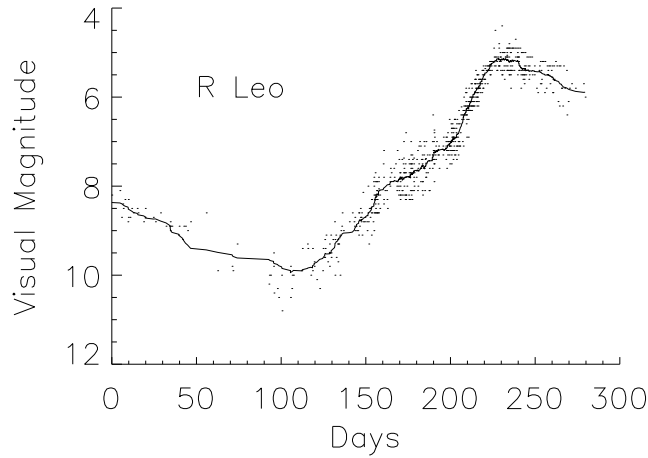,width=10.0cm}}
\caption{Visual estimates of the brightness of R Leonis over the period
analysed in this study, along with a curve generated by top-hat smoothing.}
\label{rleopub}
\end{figure}

R Leonis will serve as an example of the method of analysis.  First the
observations were combined with a smoothed curve, shown in Fig.~\ref{rleopub}.
The curve generally follows the centreline of the observations, but by
the nature of smoothing underestimates the brightness at
maximum and overestimates that at
minimum.  The two possible problem areas are kept in mind during
subsequent work. 

\begin{figure}
\centerline{\psfig{file=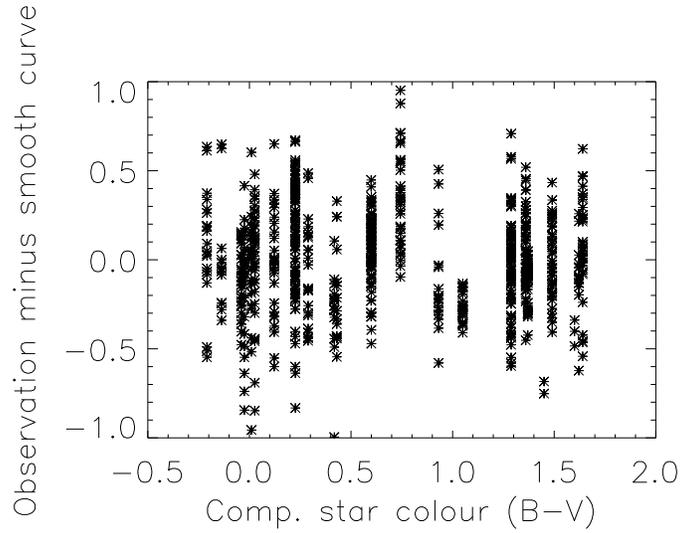,width=10.0cm}}
\caption{Deviations of visual estimates of R Leonis about the smoothed
curve, as a function of the $B-V$ colour of the closest comparison
stars.  The systematically high residuals at 0.743 are due to the
failure of the smoothed curve to trace the maximum accurately, as
the systematically low residuals at 1.05 come from the minimum.}
\label{rleo1}
\end{figure}

Next, the difference in magnitude between each observation and the smooth
curve is plotted against the colour of the closest comparison star (in
brightness and in the sky, where two comparison
stars are listed as the same brightness).  Here in Fig.~\ref{rleo1} we
see at $B-V \sim 0.743$ a clear displacement toward positive differences,
a result of the smoothed-curve error at maximum, and at $B-V \sim 1.05$
a displacement toward negative differences from a similar effect
at the minimum.  It is not
clear from this plot that there is any systematic difference between the
estimates based on very blue comparison stars (left of the plot) and
those based on red stars (on the right), though the overlap of plotting
symbols can hide a great deal. 

\begin{figure}
\centerline{\psfig{file=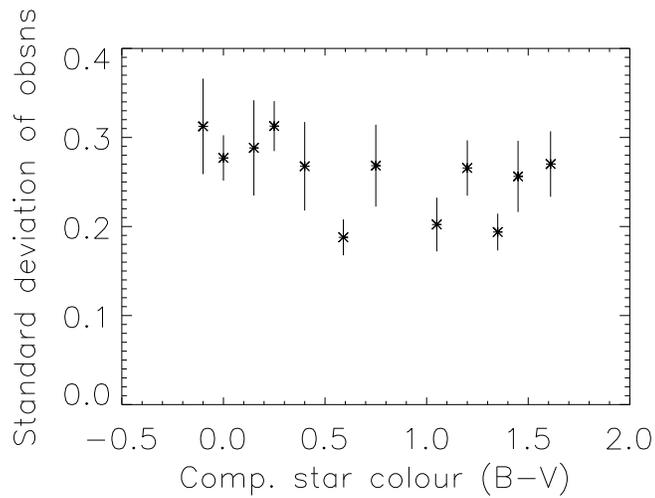,width=10.0cm}}
\caption{Standard deviation of observations of R Leonis about the average,
plotted against the colour of the closest comparison star.  Error bars are
assigned based on the number of observations in each bin.}
\label{rleo3}
\end{figure}

For that reason the data were binned (in the obvious bins, though at times
neighboring colours were combined in order to have enough observations)
and the standard deviations calculated about the mean, which gets rid of
the systematic offsets at maximum and minimum.  The resulting plot of
standard deviation against colour appears in Fig.~\ref{rleo3}.  Formal error
bars  are calculated as $\Delta \sigma = \sigma / \sqrt(n)$, with
$n$ being the number of observations in the bin,
and are probably optimistic as a measure of actual uncertainty.
The outstanding feature of this plot is its featurelessness: there is
no clear trend of dispersion of estimates with colour. (A slight trend
downward to the right can be imagined, but it is not significant.)

\begin{figure}
\centerline{\psfig{file=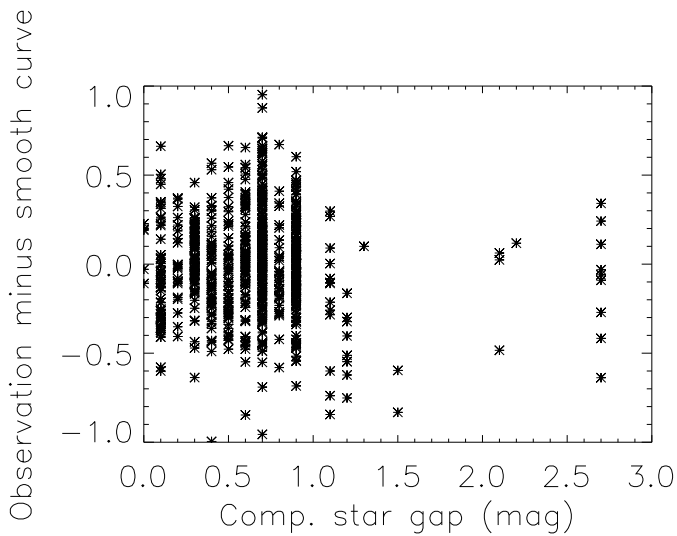,width=10.0cm}}
\caption{Differences between observations of R Leonis and a smoothed curve
plotted against the magnitude gap between the comparison stars.}
\label{rleo2}
\end{figure}

\begin{figure}
\centerline{\psfig{file=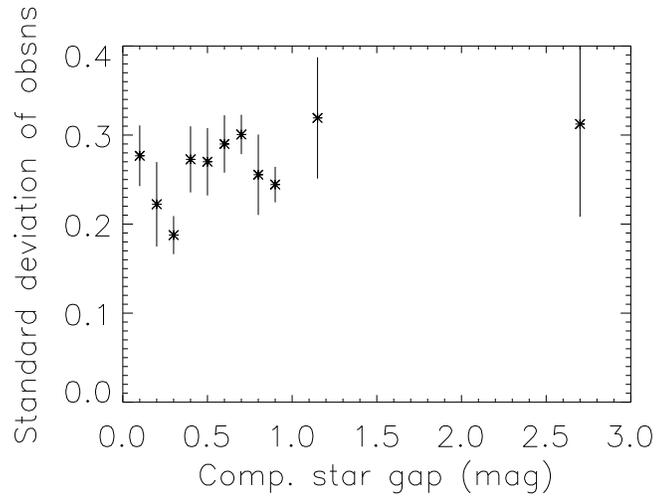,width=10.0cm}}
\caption{Standard deviation of the difference between observations and
smoothed observations of R Leonis in bins based on the magnitude gap
between comparison stars.  As before, error 
bars are based on the number of observations
in each bin.}
\label{rleo4}
\end{figure}

Postponing a discussion of this result until the data on the other variables
are presented, we turn to the question of the magnitude gap between
comparison stars.  The differences between the observations and the smoothed
curve are shown in Fig.~\ref{rleo2} plotted against the corresponding gap.
We note that there are comparatively few observations with a gap as large as
a full magnitude, some show up at 2.7.  Again, the piling up of
symbols in columns makes interpretation unclear, so as before the standard
deviations of the bins (sometimes combined) are plotted in Fig.~\ref{rleo4}.

Again we note a lack of any apparent correlation, and again we postpone
discussion until the results of all the variables have been collected.

\vspace{0.5cm}
\noindent {\it Results}

\begin{figure}
\centerline{\psfig{file=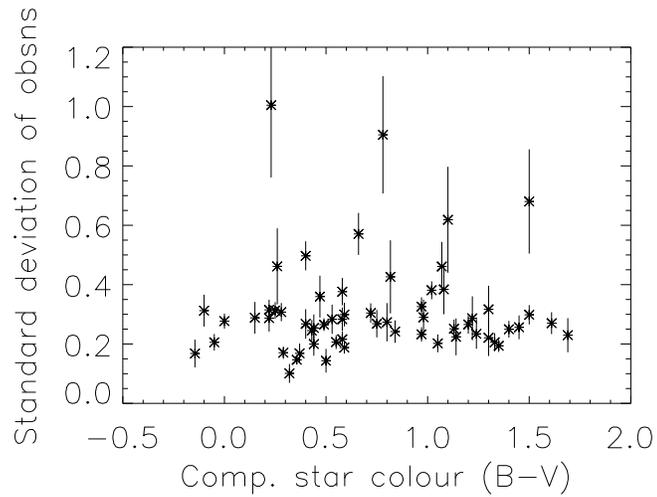,width=10.0cm}}
\caption{Standard deviation of brightness estimates for all nine
variable stars as a function of $B-V$ colour of the nearest 
comparison star.}
\label{color}
\end{figure}

\begin{figure}
\centerline{\psfig{file=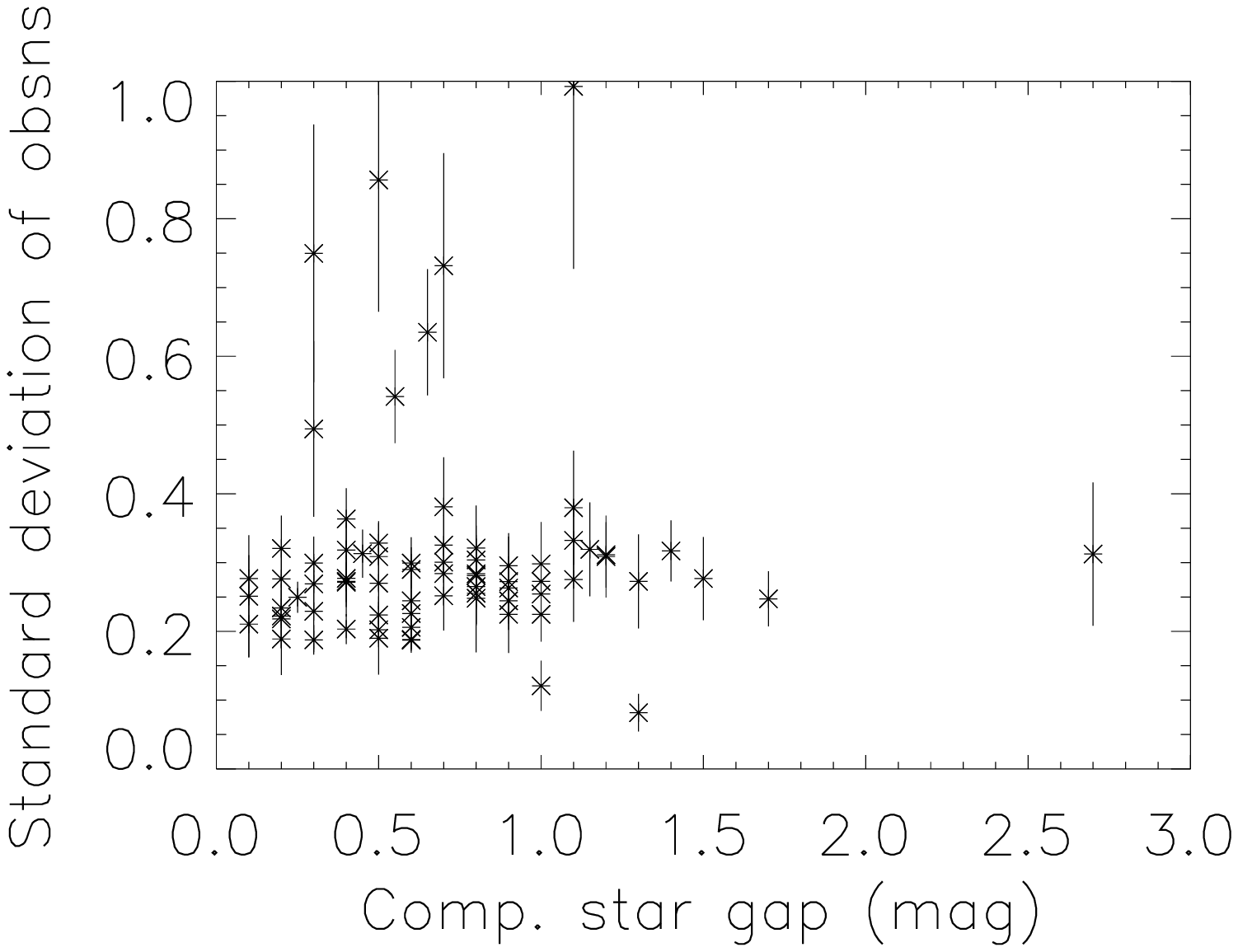,width=10.0cm}}
\caption{Standard deviation of brightness estimates for all nine
variable stars as a function of the gap between comparison star
brightness.}
\label{dcomp}
\end{figure}

The combined results of all nine stars are presented in Fig.~\ref{color}
and Fig.~\ref{dcomp}, produced as in the last section.  For several variables
the observations were much sparser than for R Leonis, resulting in fewer
bins.  In each plot there are a handful of points well above the general
trend; these (all above about $\sigma \sim 0.4$)
can be traced to an individual or a few wild points, where
an obvious mistake has been made in star identification or perhaps in
entering a Julian Date.  (Some come from the failure of a smoothed
curve to bridge a long gap between observations accurately.)
They have been left in as a reminder that
we are dealing with human data.

There is, very obviously and firmly, {\it no} trend of dispersion among the
observations with either comparison star colour or spacing of comparison
star magnitudes.  This is very
surprising.  Consider what it means.

First, with colour: for all the known problems with red stars and the
known differences among people in colour perception, it does not seem
to matter whether an $M$ Mira variable is compared with another
$M$ giant or a $B$ star; the variation in estimate among observers will
be the same.  The 0.1-magnitude disagreement in the anecdote is
dominated by some other effect, or effects, that produce a 0.2-0.3
magnitude dispersion quite reliably.

This result appears to be in flat contradiction to that of Price et al.$^4$,
who found a strong difference in the standard deviation of visual estimates
among stars of various
spectral classes.  But their work classified by the colour of the
{\it variable}, not the comparison stars, so we are not doing the same thing.
(There are several other differences between their treatment and this
one, making any detailed comparison impossible here.)



Second, it makes no apparent difference to the dispersion of estimates
among observers how far apart the comparison stars are in brightness.
This is very surprising to an observer.  One certainly has a {\it feeling}
of being on firmer ground when placing one's variable on a stepladder
of several stars 0.1-magnitude apart, rather than reaching into the
wide spaces of whole magnitudes.  But this doesn't appear to be true,
at least when comparin the estimates of several observers.  Up to a
gap of 2.7 magnitudes, visual estimates do not suffer the same 
dispersion by interpolating that they seem to do when extrapolating.

For comparison, 
consider that an
electronic detector, if limited by shot noise in the comparison star,
will have twice the uncertainty if the comparison is made 1.7 magnitudes
fainter, and three times the uncertainty for a 2.7 magnitude drop.
The situation is not, of course, directly comparable; which is
indeed the point.

Perhaps the best way to sum up these results is that the human eye-brain
system does {\it not} work like any easily-modelled detector when
performing photometry. 

\vspace{0.5cm}
\noindent {\it Implications}

It should be borne in mind that
these surprising results apply only to the observations of several or
many observers taken together; there is both anecdotal and more
systematic evidence (Skiff et al.$^4$) that observers taken
individually are significantly more reliable (with some offset)
than the 0.2-0.3 magnitude dispersion found here.

But consider the implications.  Given a ladder of stars reliably
measured as magnitude 8.1, 8.2, 8.3 and 8.4, a set of observers
will put them in any order with roughly equal probability.  More to
the point here, an observer can estimate a variable as being
{\it simultaneously} brighter than a 9.1 comparison star and fainter
than a 9.2 comparison star.  Not only will this dent the confidence
of a new observer, it can puzzle an experienced one: what number
should be reported?

On the other hand, the dispersion appears to be immune to the
obvious problems one might expect from inconvenient comparison stars.
Perhaps this result will encourage more observations of variables
now regarded as difficult and under-observed!

A deeper matter is the source of the dispersion.  Where does it come
from, and how does it behave?  There is a great deal of work yet to
be done on visual photometry.

\vspace{1.0cm}
\begin{center}
{\it Acknowledgements}
\end{center}

This study relies on the observations of many dozen AAVSO volunteer
observers, whose dedication is gratefully acknowledged though it is
impractical to list them here.  Data are \copyright 2011 by the American
Association of Variable Star Observers (AAVSO), 49 Bay State Rd., 
Cambridge, Massachusetts 02138 (U. S. A.).  All rights reserved.
No part of these data may be reproduced, transmitted, distributed, 
published, stored in an information retrieval system, posted to any
online or ftp site, or otherwise communicated, in printed form or
electronically, without the express written permission of the AAVSO.

\vspace{1.0cm}
\begin{center}
{\it References}
\end{center}

\noindent (1) John C. Martin, Kris Davidson, Roberta M. Humphreys \& Andrea
Mehner, {\em AJ}, {\bf 139}, 2056, 2010 \\

\noindent (2) Rev. T. W. Webb, {\em Celestial Objects for Common Telescopes,
Vol. 2}, Dover Edition edited and revised by Margaret W. Mayall (Dover,
New York) 1962 (original edition 1859), p. 7 \\

\noindent (3) A. A. Henden, 2011, Observations from the AAVSO International
Database, private communication.

\noindent (4) A. Price, G. Foster \& B. Skiff, {\em The Precision of
Visual Estimates of Variable Stars}, poster at AAS meeting, 2007 January

\end{document}